\documentclass[preprint,12pt,english]{elsarticle}
\usepackage{amsmath}
\usepackage{amssymb}
\usepackage{lipsum}
\usepackage[sans]{dsfont}
\usepackage{multirow}
\usepackage{afterpage}
\usepackage{amsfonts}
\usepackage{bm}%
\usepackage{babel}
\usepackage{hyperref}
\usepackage{graphicx}
\usepackage{color}
\def\fnum@figure{\figurename\thefigure}
\renewcommand{\figurename}{Fig.}

\journal{Physica A}

\title{Entanglement and Measurement-induced quantum correlation in Heisenberg spin models}
\begin{document}
\author{Indrajith V S$^*$ , R. Muthuganesan$^\ddag$ R. Sankaranarayanan$^*$ }
\address{$^*$ Department of Physics, National Institute of Technology\\ Tiruchirappalli-620 015, Tamil Nadu, India.}
\address{$^\ddag$ Centre for Nonlinear Science and Engineering, School of Electrical and Electronics Engineering,  SASTRA Deemed University, Thanjavur - 613 401, Tamil Nadu, India}
\begin{abstract}
Correlation beyond entanglement is a subject of interest in quantum information. Here we showed the existence of quantum correlation without entanglement in Heisenberg $XYZ$ spin model with external magnetic field, using different versions of measurement-induced nonlocality. Maximally entangled states are shown to possess maximum correlation.

\end{abstract}


\maketitle

\section{Introduction}
Nonlocality refers to the weird correlation between different parts of composite system in quantum domain \cite{corltn}. Within the ambit of local hidden variable theory, violation of Bell inquality serves as a test of nonlocality of a quantum state \cite{bel}. Since all entangled pure states violate Bell inquality, entanglement is viewed as a manifestation of nonlocality. This makes entanglement a useful resource in various information processing including teleportaion \cite{telepor}, dense coding \cite{dens_cod} etc. However, Werner showed that there exist unentangled mixed states which violate Bell inequality \cite{werner}. In this sense, entanglement is not sufficient for complete manifestation of nonlocality. This observation necessitates to look for a broader framework of nonlocality, which must be beyond the entanglement of states.   

From the seminal work of Oillivier and Zurek, it is clear that all nonlocally correlated states need not be entangled \cite{discord}. They quantified quantum correlation interms of mutual information called quantum discord. Such correlation is shown to exist even for separable (unentangled) states. Subsequently, it is shown that almost all quantum states have non vanishing discord \cite{non_van_dis}. Further development on this direction showed the necessary and sufficient condition for non-vanishing quantum discord, leading to a geometric version of discord which is also experimentally accessible \cite{geo_discord}. 

In 2011 S. Luo and S. Fuo proposed another correlation measure in geometric perspective based on projective measurements, namely measurement-induced nonlocality (MIN) \cite{min}, which is a dual to the geometric discord. It is worth noting that many quantum operations such as superdense coding, teleportation etc. involve local measurements and comparing pre-- and post-- measurement states. However, MIN encounters the so called local ancilla problem \cite{piani}, in which the quantity may change arbitrarily due to an unmeasured ancilla state. This problem can be tackled by replacing density matrix with its square root \cite{tracesqroot}. In recent years, MIN has also been investigated based on relative entropy \cite{relativemin}, skew information \cite{minskew}, trace distance \cite{tmin} and fidelity \cite{fmin} to resolve the local ancilla problem. A good number of works on non-local aspects of two qubit system with different motivations can be found elsewhere  \cite{review_11,review_12,review_13,review_14}.

In this article we investigate the nonlocal correlation of bipartite state of simple physical system, namely Heisenberg spin model in thermal equlibrium. The correlation is quantified using various forms of MIN and compared with spin entanglement. The comparison is made systematically with respect to various system parameters. In particular, nonlocal correlation between the spins in the absence of entanglement is brought out with sufficient illustations.
%%%%%%%%%%%%%%%%%%%%%%%%%%%%%%%%%%%%%%%%%%%%%%%%%%%%%%%%%%%%%%%%%%%%%%%%%%%%%%%%%%%%%%%%%%%%%%%%%%%%%%%%
\section{Preliminaries}
\subsection{Entanglement}

Hereafter the entanglement between subsystems is measured using concurrence. For a bipartite state $\rho$ shared by the parties $a$ and $b$  with corresponding Hilbert spaces $\mathcal{H}^a$ and $\mathcal{H}^b$ respectively, the concurrence is defined as \cite{wotters97}
\begin{equation}
C(\rho)=\text{max}\lbrace0,\sqrt{\lambda_1}-\sqrt{\lambda_2}-\sqrt{\lambda_3}-\sqrt{\lambda_4}\rbrace 
\end{equation}
where $\lambda_i $ are eigenvalues of matrix $\rho\tilde{\rho}$ arranged in decreasing order and $\tilde{\rho}=(\sigma_y\otimes\sigma_y)\rho^{*}(\sigma_y\otimes\sigma_y)$  is spin flipped matrix. The concurrence lies between $0$ and $1$, such that minimum and maximum values correspond to separable (unentangled) and maximally entangled states respectively.
\subsection{Measurement-Induced Nonlocality (MIN)}
It is a correlation measure in the geometric perspective to capture nonlocal effect on quantum state due to local projective measurements. This quantity in some sense is dual to geometric quantum discord \cite{geo_discord}, and is defined as \cite{min}
\begin{equation}
N_2(\rho):=~ ^{\text{max}}_{\Pi^a}\Vert\rho-\Pi^a(\rho)\Vert^2
\end{equation}
where $\Vert \mathcal{A} \Vert = \sqrt{\text{Tr}(\mathcal{A}^{\dagger}\mathcal{A})}$ is the Hilbert-Schmidt norm of the operator $\mathcal{A}$. Here the maximum is taken over all possible von Neumann projective measurements $\Pi^a=\lbrace\Pi^a_k\rbrace = \lbrace\vert k \rangle \langle k \vert \rbrace ~\text{and}\,\Pi^a(\rho) = \sum_k(\Pi_k^a \otimes \textbf{I}^b)\rho(\Pi_k^a \otimes \textbf{I}^b)$.

An arbitrary state of a bipartite $m \times n$ dimensional composite system can be written as
\begin{equation}
\rho= \sum_{i,j}\gamma _{ij}X_{i}\otimes  Y_{j} \label{state1}
\end{equation}
where $\Gamma = (\gamma _{ij} =\hbox{Tr} (\rho\,X_{i}\otimes Y_{j}))$ is an $m^2 \times n^2$ real matrix, $X_i$
and $Y_j$ are orthonormal operators for subsystem $a$ and $b$ respectively. If $X_{0}=\mathbf{I}^{a}/\sqrt{m}$, $Y_{0}=\mathbf{I}^{b}/\sqrt{n}$, and separating the terms in eq.(\ref{state1}) the state $\rho$ can be written as  
\begin{equation}
\rho =\frac{1}{\sqrt{m n}}\frac{\mathbf{I}^{a}}{\sqrt{m}}\otimes \frac{\mathbf{I}^{b}}{\sqrt{n}}+\sum_{i=1}^{m^2 -1}x_{i}X_{i}\otimes\frac{\mathbf{I}^{b}}{\sqrt{n}}+\frac{\mathbf{I}^{a}}{\sqrt{m}}\otimes\sum_{j=1}^{n^2 -1}y_{j}Y_{j} +\sum_{i,j\neq 0}t_{ij }X_{i}\otimes Y_{j} \label{state2}
\end{equation}
where   $x_{i}=\hbox{Tr}(\rho\,X_{i}\otimes \mathbf{I}^{b})/\sqrt{n}$, $y_{j}= \hbox{Tr}(\rho\,\mathbf{I}^{a} \otimes Y_{j} )/\sqrt{m}$ and $T = (t _{ij} = \hbox{Tr} (\rho\,X_{i}\otimes  Y_{j}))$ is a real correlation matrix of order $(m^{2}-1)\times(n^2 -1)$.

In fact, MIN has closed formula for  $2\times n$ dimensional system as
\begin{equation}
N_{2}(\rho)= \begin{cases}
\text{Tr}(TT^t)-\frac{1}{\Vert \textbf{x} \Vert^2}\textbf{x}^tTT^t\textbf{x} &  \text{if} \quad \textbf{x}\neq 0, \\
\text{Tr}(TT^t)-\lambda_{\text{min}} &  \text{if} \quad \textbf{x}=0, 
\end{cases}
\end{equation}
where $\lambda_{\text{min}}$ is the least eigenvalue of $3\times 3$ dimensional matrix $TT^t$ and $\textbf{x} = (x_1~x_2~ x_3) $. 
\subsection{Trace MIN (T-MIN)}

Due to easy computation and experimental realization \cite{exp_real}, much attention has been paid on MIN in recent years. However, this quantity is not a bonafide measure of quantum correalation as it suffers from local ancilla problem \cite{piani}. One alternate form of MIN based on trace distance, namely trace MIN (T-MIN) resolves this problem \cite{tmin}. It is defined as
\begin{equation}
N_1(\rho):= ~^{\text{max}}_{\Pi^a}\Vert\rho-\Pi^a(\rho)\Vert_1
\end{equation} 
where $\Vert \mathcal{A} \Vert_1 = \text{Tr}\sqrt{\mathcal{A}^{\dagger}\mathcal{A}}$ is the trace norm of operator $\mathcal{A}$. Here also the maximum is taken over all von Neumann projective measurements.
Rewriting two qubit state (\ref{state2}) as 
\begin{equation}
\rho=\frac{1}{4}\bigg(\mathbf{I}\otimes\mathbf{I}+\textbf{x} \cdot \mathbf{\sigma}\otimes\mathbf{I}+\mathbf{I}\otimes\textbf{y} \cdot \sigma+\sum^3_{i=1}c_i\sigma_i\otimes\sigma_i\bigg)
\end{equation}
where $\mathbf{\sigma} = (\sigma_1~\sigma_2~\sigma_3)$, $\textbf{y} = (y_1~y_2~y_3)~\text{and}~\textbf{c} =  (c_1~c_2~c_3) ~ \text{with} ~ c_i = \text{Tr}(\rho\,\sigma_i \otimes \sigma_i)$,
the closed formula of $N_1(\rho)$ is given as 
\begin{equation}
N_1(\rho)=
\begin{cases}
\frac{\sqrt{\chi_+}~+~\sqrt{\chi_-}}{2 \Vert \textbf{x} \Vert_1} & 
 \text{if} \quad \textbf{x}\neq 0,\\
\text{max} \lbrace \vert c_1\vert,\vert c_2\vert,\vert c_3\vert\rbrace &  \text{if} \quad \textbf{x}=0,
\end{cases}
\end{equation}
where $\chi_\pm~=~ \alpha \pm 2 \sqrt{\tilde{\beta}} \Vert \textbf{x} \Vert_1 ,\alpha =\Vert \textbf{c} \Vert^2_1 ~\Vert \textbf{x} \Vert^2_1-\sum_i c^2_i x^2_i,\tilde{\beta}=\sum_{\langle ijk \rangle} x^2_ic^2_jc^2_k, \vert c_i \vert $ is the absoulte value of $c_i$ and the summation runs over cyclic permutation of $\lbrace 1,2,3 \rbrace$.
\subsection{ Fidelity MIN (F-MIN) }
Since fidelity itself is not a metric, any monotonically decreasing function of fidelity defines a valid distance measure. Defining MIN based on fidelity induced metric as \cite{fmin}
\begin{equation}
N_{\mathcal{F}}(\rho)=~1-~^{\text{min}}_{\Pi^a}\mathcal{F}(\rho,\Pi^a(\rho))
\end{equation}
where $ \mathcal{F}$ is the fidelity between the states $\rho$ and $\sigma $ defined as \cite{wang}
\begin{equation}
 \mathcal{F} =\frac{(\text{Tr}(\rho\sigma))^2}{\text{Tr}(\rho)^2 \text{Tr}(\sigma)^2} \nonumber
\end{equation}
which satisfies axioms of original fidelity \cite{jozza}.

The closed formula of Fidelity based MIN (F-MIN) for $2 \times n$ dimensional system is given as \cite{fmin}
\begin{equation}
N_{\mathcal{F}}(\rho)=  
\begin{cases}
\frac{1}{\| \Gamma  \|^{2}}(\| \Gamma  \|^{2}-\mu _{1}) & \text{if}~~~ \textbf{x}=0 ,\\ \frac{1}{\| \Gamma  \|^{2}}(\| \Gamma  \|^{2}-\epsilon ) & \text{if}~~~\textbf{x}\neq0,
\end{cases}
\end{equation}
where $\epsilon =\hbox{Tr}(A\Gamma \Gamma ^{t}A^{t})$ and $\mu_1$ is the smallest eigenvalue of $\Gamma\Gamma^t$ and
\begin{equation} \label{eq:ope}
A=\frac{1}{\sqrt{2}}
\begin{pmatrix}
1 & \frac{\textbf{x}}{\| \textbf{x} \|}\\
1 & -\frac{\textbf{x}}{\| \textbf{x} \|}
\end{pmatrix}.
\end{equation}
%%%%%%%%%%%%%%%%%%%%%%%%%%%%%%%%%%%%%%%%%%%%%%%%%%%%%%%%%%%%%%%%%%%%%%%%%%%%%%%%%%%%%%%%%%%%%%%%%%%%%%%%%%%%%
\section{The Hamiltonian}
In order to study the behaviour of different measures of quantum correlation along with entanglement, we consider the scaled dimensionless Hamiltonian of two spin  - $\frac{1}{2} $ system as\\ 
 \begin{equation}
 H = \frac{J}{2}[(1+\gamma)\sigma_{x}^{1}\sigma_{x}^{2}+(1-\gamma)\sigma_{y}^{1}\sigma_{y}^{2}]+  \frac{1}{2}[J_{z}\sigma_{z}^{1}\sigma_{z}^{2}+(B+\lambda)\sigma_{z}^{1}+(B-\lambda)\sigma_{z}^{2}] \label{Hamiltonian}
 \end{equation}\\
 where $\sigma_{k} $ are the pauli spin matrices, $\gamma = (J_{x}-J_{y})/(J_{x}+J_{y})$ is the anisotropy in $XY$ plane with $J_{k}$ being the interaction strength in respective spin components, $B$ is  the strength of magnetic field and $\lambda$ signifies inhomogeneity in the field. The matrix form of the Hamiltonian (\ref{Hamiltonian}) in standard two qubit computational basis is given as
\begin{equation}
H=
\begin{pmatrix}
\frac{J_z}{2}+B & 0 & 0 & \gamma J\\
0 & -\frac{J_{z}}{2}+\lambda & J & 0\\
0 & J & -\frac{J_{z}}{2}-\lambda & 0\\
\gamma J & 0 & 0 & \frac{J_z}{2}-B
\end{pmatrix}.
\end{equation}
The above Hamiltonian  is diagonalized and the corresponding eigenvalues  and eigenvectors are

\begin{equation}
E_{1,2}=\frac{J_{z}}{2}\pm \eta\hspace{3cm}\vert\psi_{1,2}\rangle=N_{\pm }\bigg(\frac{B\pm \eta}{\gamma J}\vert00\rangle+\vert11\rangle\bigg)
$$
$$
E_{3,4}=-\frac{J_{z}}{2}\pm \delta\hspace{3cm}\vert\psi_{3,4}\rangle=M_{\pm }\bigg(\frac{\lambda\pm \delta}{J}\vert01\rangle+\vert10\rangle\bigg) \nonumber
\end{equation}
where, $\delta=\sqrt{{\lambda}^2+ J^2}$, ~$\eta=\sqrt{B^2+(\gamma J)^2}$ and the normalisation constants $N_{\pm} =((B\pm\eta/\gamma J)^2+1)^{-1/2}$, $M_{\pm} =((\lambda\pm\delta/J)^2+1)^{-1/2}$. For $B=\lambda =0$ and $\gamma = 0,~ J = J_{z}$ the Hamiltonian corresponds to Heisenberg spin with isotropic interaction, and the eigenfunctions are reduced to maximally entangled  Bell states.

Thermal state of the above Hamiltonian is given by $\rho(T)=e^{-\beta H}/\mathcal{Z}$, where $\beta = 1/k_B T$ with $k_B$ being the Boltzmann constant, $T$ is the equilibrium temperature and $\mathcal{Z} = \text{Tr} (e^{-\beta H})$  is the partition function. Setting $\beta=1$, implying that the energy is scaled such that $k_B T = 1$, the thermal state in computational basis is obtained as

 \begin{equation}
 \rho = \frac{1}{\mathcal{Z}}
 \begin{pmatrix}
 \mu_{-} & 0 & 0 & \kappa\\
 0 & \nu_{-} & \epsilon & 0\\
 0 & \epsilon & \nu_{+} & 0\\
 \kappa & 0 & 0 & \mu_{+}\\ \label{Thermal}
 \end{pmatrix} 
 \end{equation}
 with matrix elements $\mu_{\pm}=e^{-\frac{J_{z}}{2}}(\cosh \eta ~{\pm}~\frac{B}{\eta}\sinh\eta)$, $\kappa=-\frac{\gamma J}{\eta}e^{\frac{-J_{z}}{2}}\sinh\eta$, $\nu_{\pm}=e^{\frac{J_{z}}{2}}(\cosh \delta~{\pm}~\frac{\lambda}{\delta}\sinh\delta)$,  $\epsilon=-\frac{J}{\delta}e^{\frac{J_{z}}{2}}\sinh\delta$ and the partition function $\mathcal{Z}=2(e^{-\frac{J_{z}}{2}}\cosh\eta+e^{\frac{J_{z}}{2}}\cosh\delta)$. 
 %%%%%%%%%%%%%%%%%%%%%%%%%%%%%%%%%%%%%%%%%%%%%%%%%%%%%%%%%%%%%%%%%%%%%%%%%%%%%%%%%%%%%%%%%%%%%%%%%%%%%%%%%%%%
 \section{Results and Discussion}
The concurrence for the above thermal state (\ref{Thermal}) is evaluated as
\begin{equation}
C(\rho)=2~\text{max}\left\lbrace 0,\frac{\vert\kappa\vert-\sqrt{\nu_{+}\nu_{-}}}{\mathcal{Z}},\frac{\vert\epsilon\vert-\sqrt{\mu_{+}\mu_{-}}}{\mathcal{Z}} \right\rbrace \, . 
\label{concurence}
\end{equation} 
From this expression, it can be shown that $C(\rho) = 0$ for $J_{c_1} \leq J_{z} \leq J_{c_2}$  where 
\begin{equation}
  J_{c_1} = \ln\Bigg(\frac{\gamma J \sinh\eta}{\eta\cosh J}\Bigg),~~J_{c_2} = \frac{1}{2}\ln\Bigg(\frac{\cosh^2\eta-\frac{B^2}{\eta^2}\sinh^2\eta}{\sinh^2J}\Bigg).
  \label{critical pnts}
\end{equation}

Further the measurement-induced quantum correlation for the above thermal state is quantified using various forms of MINs. The MIN, T-MIN and F-MIN are computed as 
\begin{equation}
N_2(\rho) = \frac{2(\kappa^2 + \epsilon^2)}{\mathcal{Z}^2},
\end{equation}
 
\begin{equation}
N_{1}(\rho) = \frac{\vert\kappa\vert + \vert\epsilon\vert}{\mathcal{Z}},
\end{equation}

\begin{equation}
N_{\mathcal{F}}(\rho) = \frac{2(\kappa^2 + \epsilon^2)}{2(\kappa^2 + \epsilon^2)+(\mu_{-}^2 + \nu_{+}^2 + \nu_{-}^2 + \mu_{+}^2)} \, .
\end{equation} 
 Since the value of MINs range from 0 to 0.5, we have considered $C(\rho)/2$ for better comparison. From the above results we observe that all the form of MINs are identically zero, only if $\kappa= \epsilon = 0$. Hence the interaction strength $J$ is significant for the measurement induced quantum correlation between the spins.
\subsection{$XXX$ and $XXZ$ Model}
To study the above results in detail we first set $B = \lambda = 0$. We also set the spin interaction such that $\gamma = 0$ for which $J_{c_1} = -\infty$ and $J_{c_2} = - \ln (\sinh J)$. With this setting we consider the following two cases namely (i)$J = J_z$, $XXX$ model and (ii) $J \neq J_z$, $XXZ$ model. 

\begin{figure*}[!ht]
\centering\includegraphics[width=0.7\linewidth]{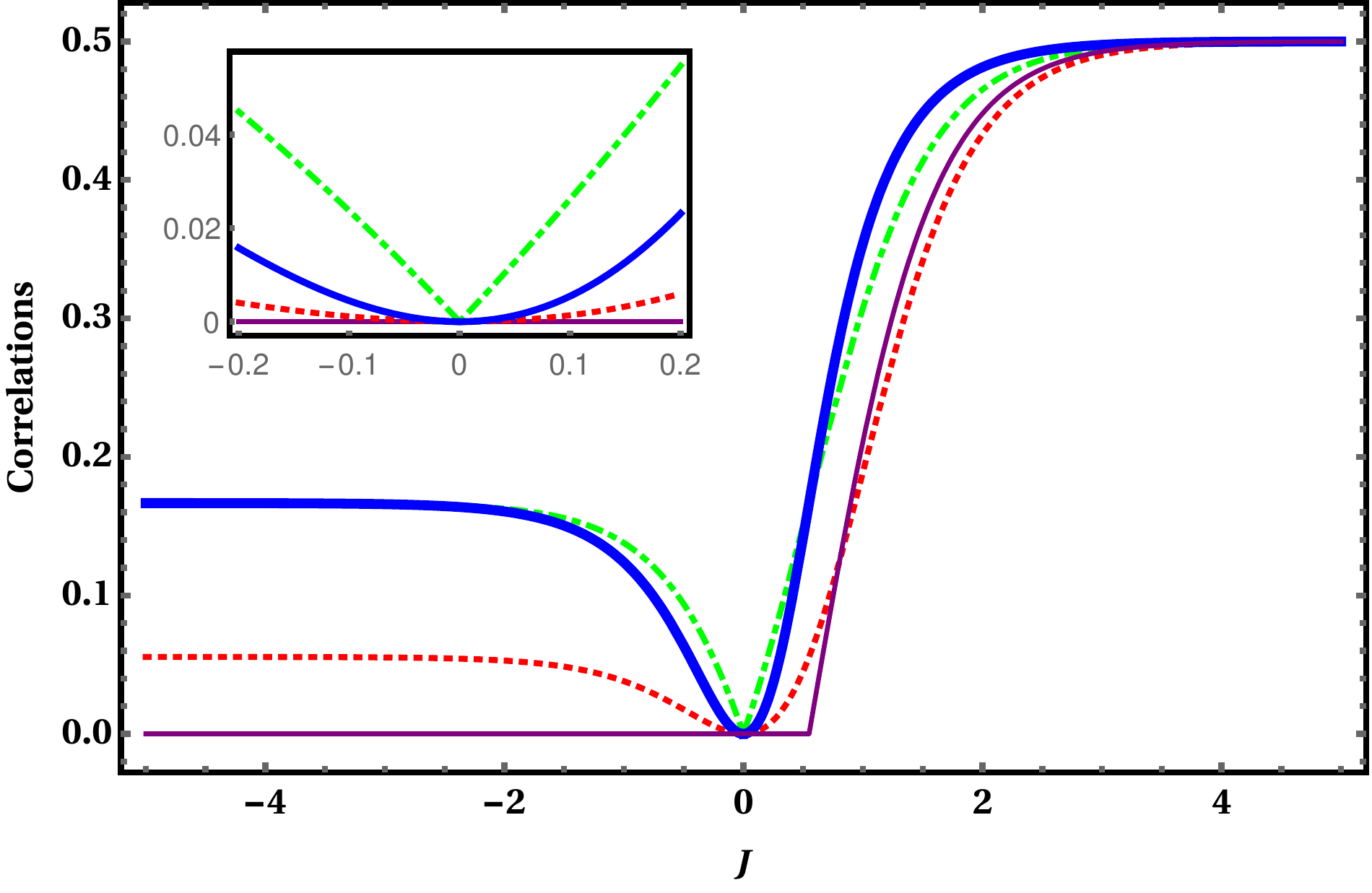}
\caption{(color online) Concurrence (thin), MIN (dotted), T-MIN (dot-dashed) and F-MIN (bold) for XXX model with $B = \lambda = 0$. Inset shows the maginification in the neighbourhood of $J = 0$.}
\label{fig13}
\end{figure*}

(i) For $XXX$ model, the critical value $J_{c_2} = \ln{3}/2$ and thus entanglement between the spins are absent for $J_z \leq  \ln{3}/2$. On the other hand MIN, T-MIN and F-MIN are zero only if $J_z $  or $J = 0$, as mentioned earlier. In  Fig. \ref{fig13}, we plotted all the MINs along with the entanglement. We observe from the figure that all the MINs have similar behavior. For $J \neq 0$, $\kappa \neq 0~\text{and} ~\epsilon \neq 0$ implying the off-diagonal elements of the state $\rho$ are non zero, which inturn induce coherence. These coherence terms are indeed reflecting the quantum correlation between the spins. Thus it is clear that all the MINs are consistent in quantifying the spin correlation. In particular, all the MINs are reflecting 
non-zero correlation between the spins even when the spins are unentangled. It is also seen that the correlation increases to the maximum value of 0.5 only in the antiferromagnetic ($J > 0$) phase. Further, the plot also confirms that if the spins are maximally entangled, quantum correlation measured by all the MINs are also maximum. The corresponding states could be a mixture of maximally entangled states \cite{twosided}.

\begin{figure*}[!ht]
\includegraphics[width=0.5\textwidth]{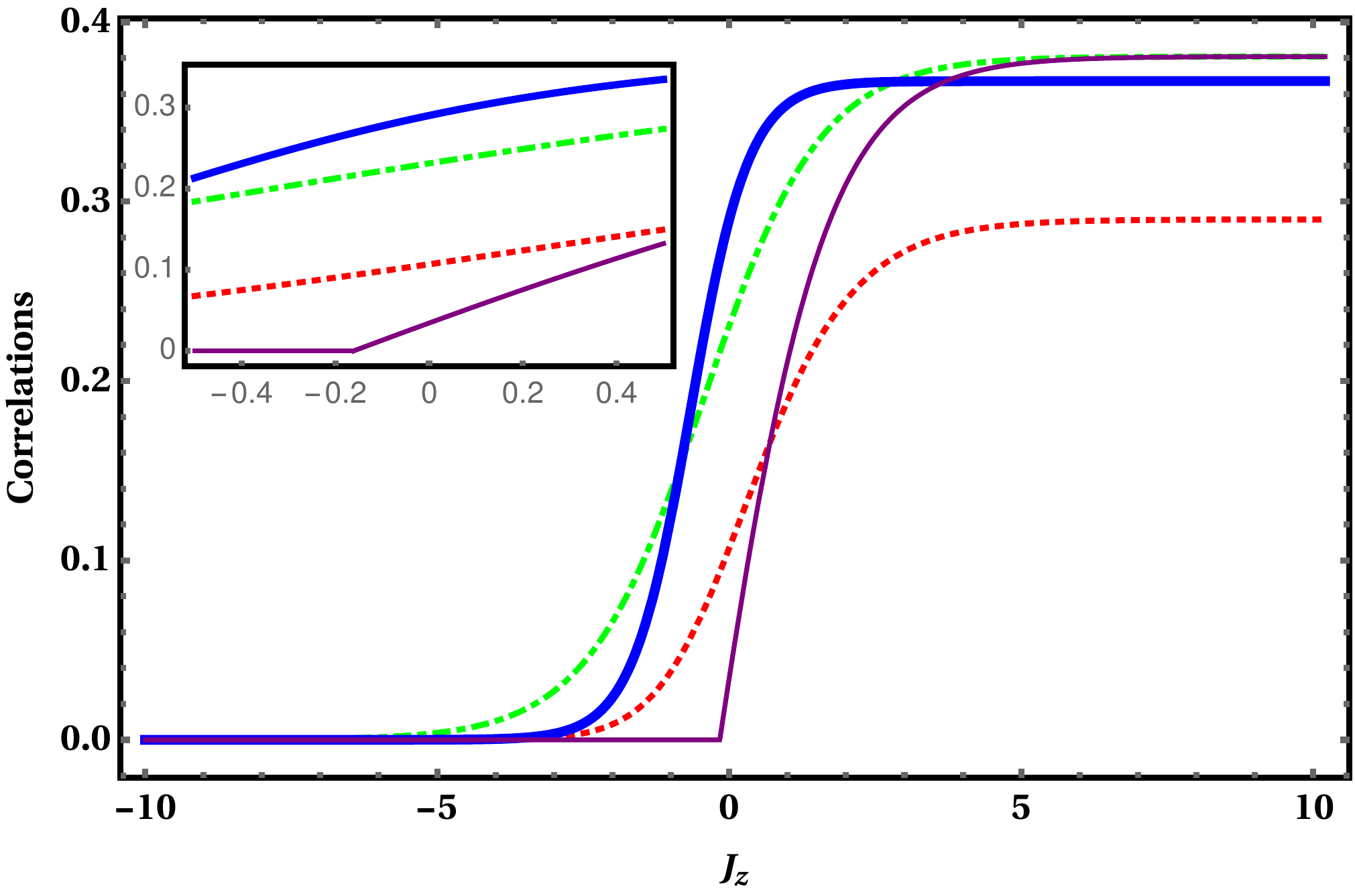}
\includegraphics[width=0.5\textwidth]{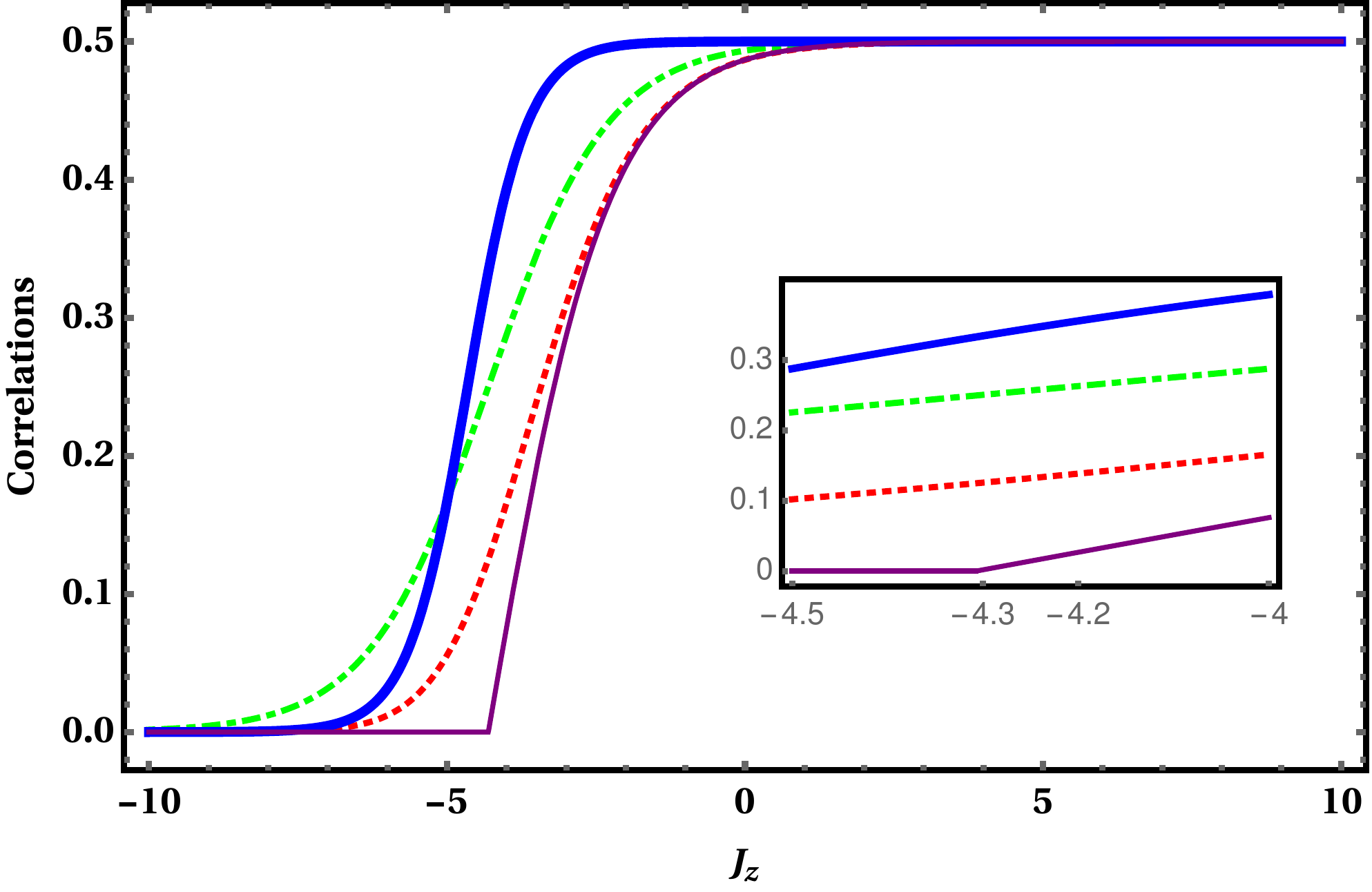}
\caption{(color online) Setting $B = \lambda = 0$, Concurrence (thin), MIN (dotted), T-MIN (dot-dashed) and F-MIN (bold) for XXZ model with $J = 1 $ (left) and $J = 5$ (right). Insets show the magnification around $J_{c_2}$.}
\label{figxxz}
\end{figure*}

(ii) For $XXZ$ model ($J \neq J_z$), the results are plotted in Fig. \ref{figxxz}, where we observe that both entanglement and MINs are significantly enhanced with $J$ at $J_z = 0$. This imply that spin entanglement and correlation are increasing with interaction in $XX$ model. Further, the behavior of MINs are somewhat similar to that of entanglement, except that MINs are smooth functions of the parameter $J_z$. Here also a part of the region $(J_z < J_{c_2})$ where spins possess correlation without being entangled is visible.
 If the interaction of spins in $xy$-plane ($J$) is sufficiently large, the region of maximum entanglement also exhibits maximum correlation as measured by different MINs. The critical points $J_{c_2} = - \ln (\sinh J)$ are calculated as -0.161 and -4.306 for $J = 1 $ and $J = 5 $ respectively, which are clearly visible in the inset plots.

Having studied the role of interaction strength of spins, now we look at the role of magnetic field on entanglement and correlation. A typical plot with strength of magnetic field $B$ is shown in Fig. 3. Here we observe that any increase of $B$ decrease both the entanglement and correlation between the spins from their maximum values 0.5. This result  shows that the two-spin system looses its quantum signature with the increase of external magnetic field. 
\begin{figure*}[!ht]
\centering\includegraphics[width=0.7\textwidth]{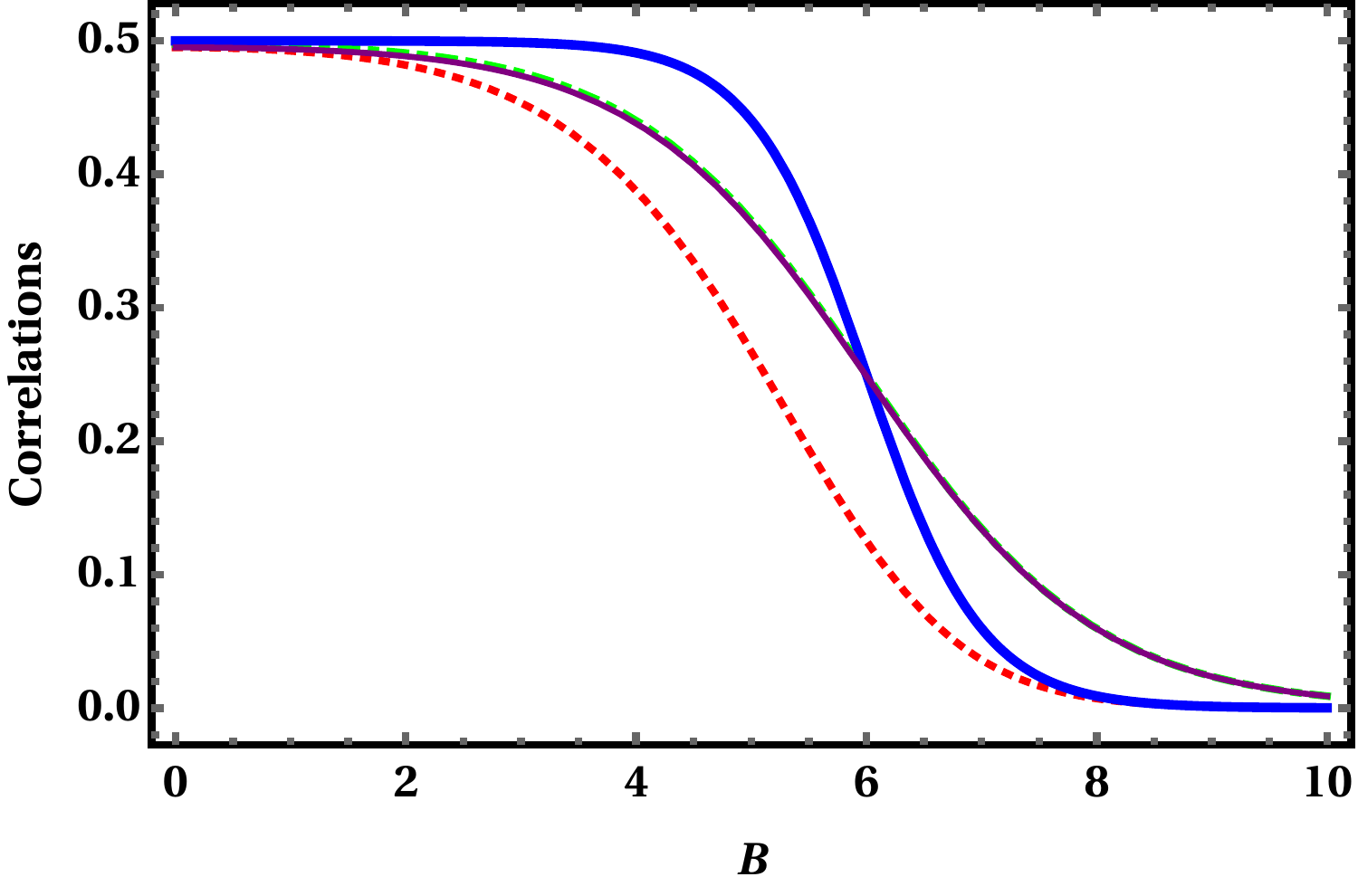}
\caption{(color online) Concurrence (thin), MIN (dotted), T-MIN (dot-dashed) and F-MIN (bold) for XXZ model. Here we set $J = 5, J_z = 1$ and $\lambda = 0$.}
\end{figure*}
\label{xxzB}

\subsection{$XYZ$ Model}  
\begin{figure*}[!ht]
\includegraphics[width=0.5\linewidth]{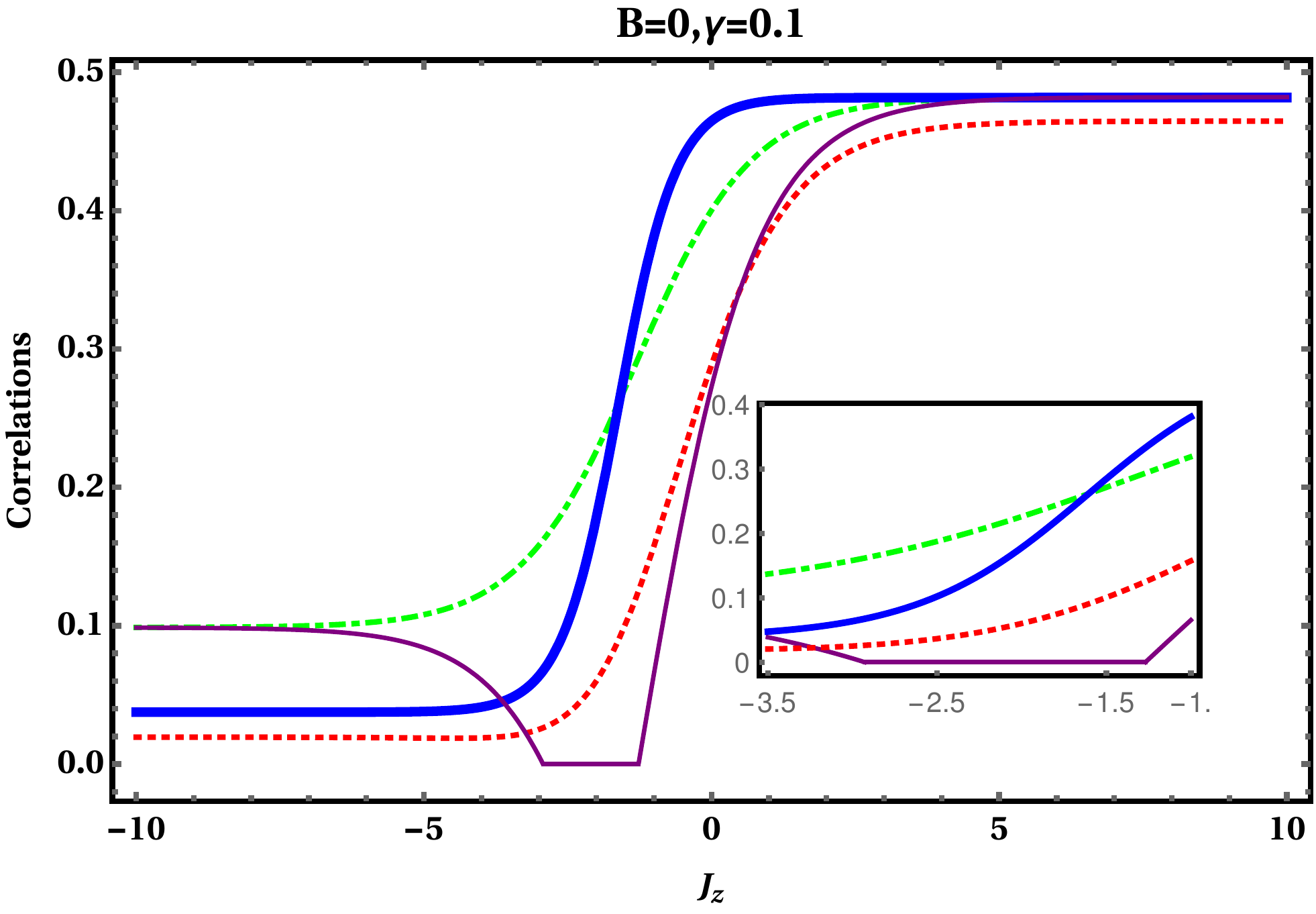}
\includegraphics[width=0.5\linewidth]{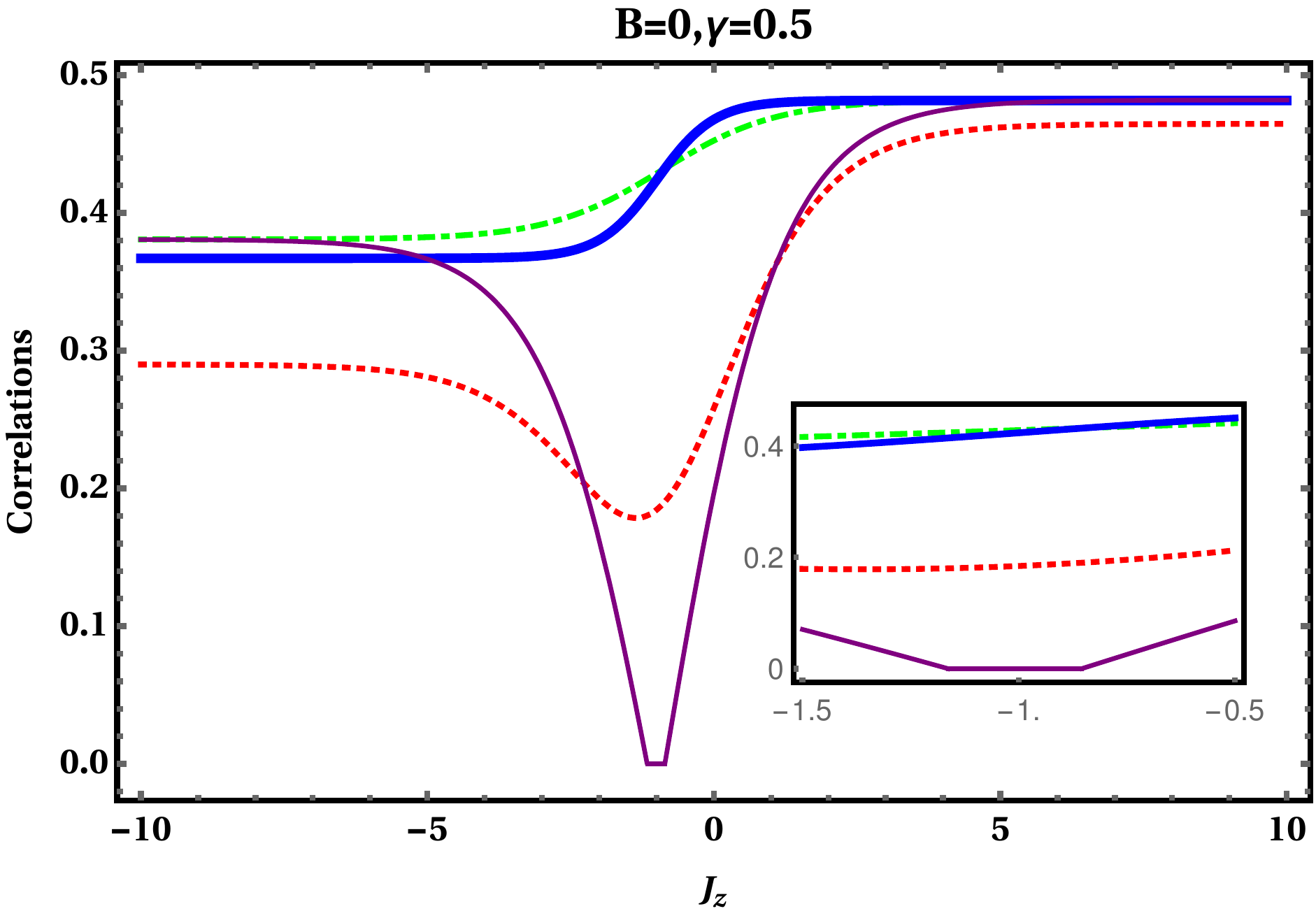}\\
\includegraphics[width=0.5\linewidth]{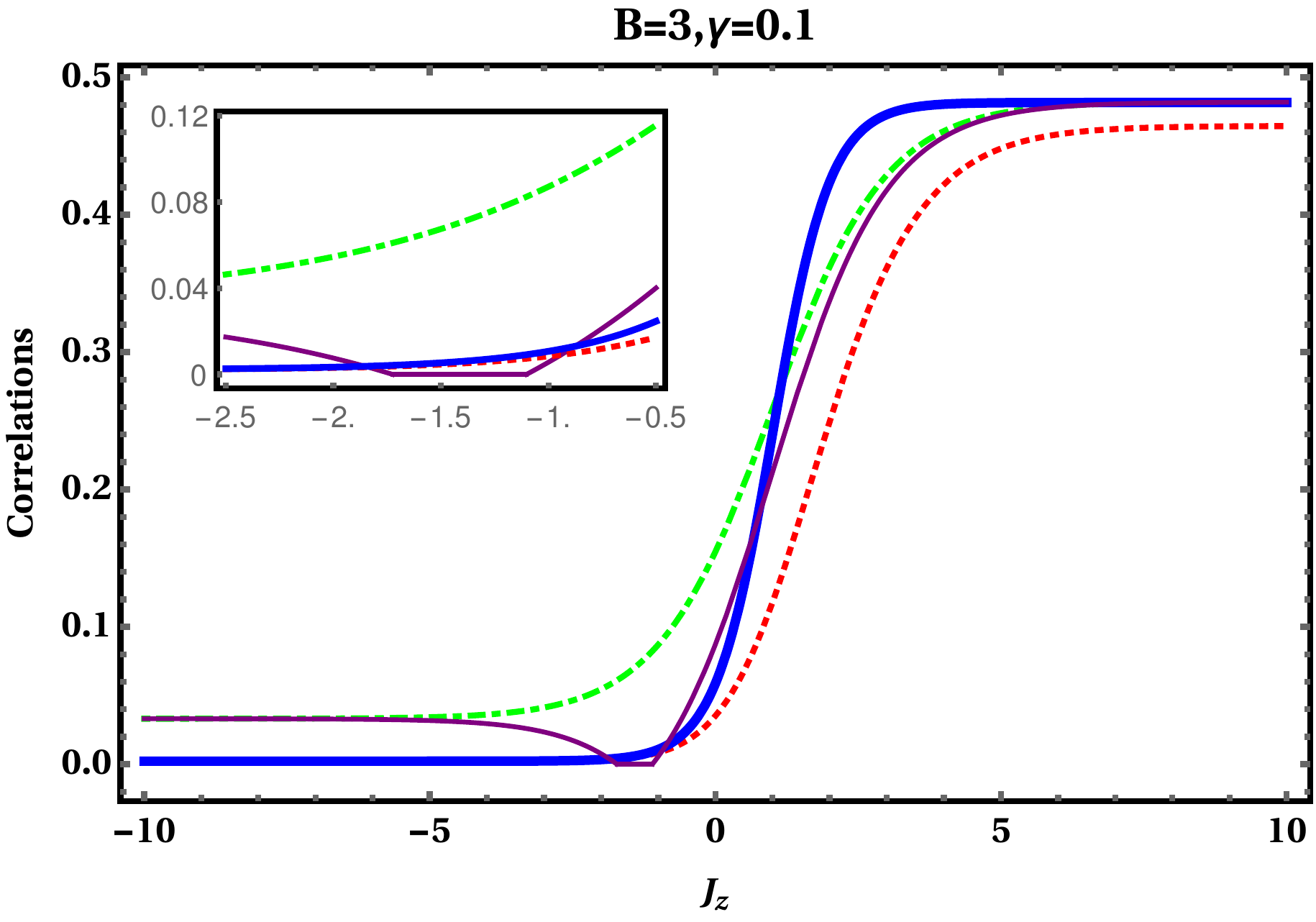}
\includegraphics[width=0.5\linewidth]{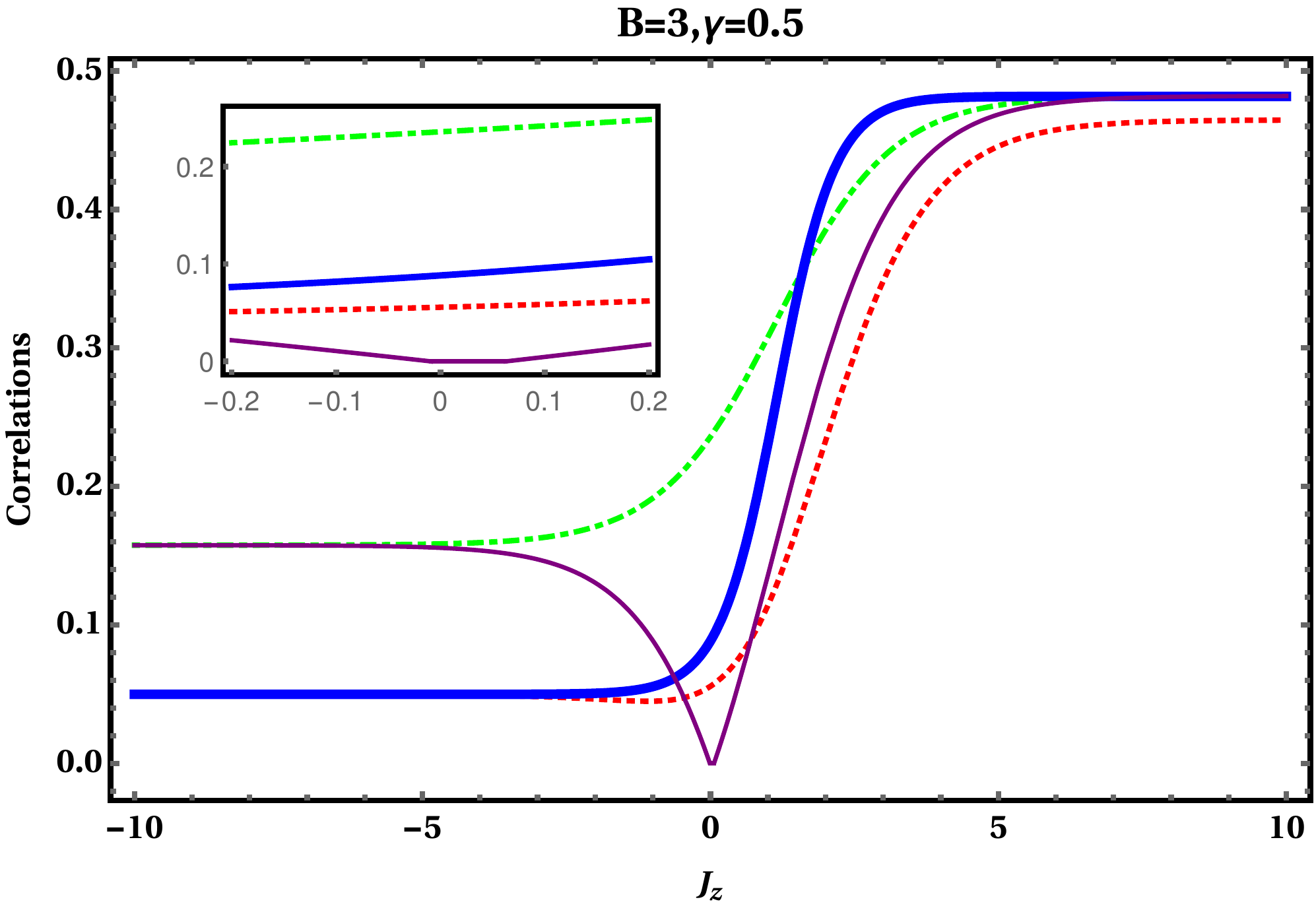}
\caption{(color online) Concurrence (thin), MIN (dotted), T-MIN (dot-dashed) and F-MIN (bold) for XYZ model with $J = 2$ and $\lambda = 0$. Insets show the values $J_{c_1}$ and $J_{c_2}$ between which the spins are unentangled.}
\label{fig}
\end{figure*}

Now we extend our analysis on general anisotropic Heisenberg spins ($\gamma \neq 0 $). Since all the quantities of our interest are even functions of anisotropic parameter $\gamma$ and magnetic field $B$, we look at them for $\gamma > 0$ and $B > 0$. Setting $\lambda = 0$, typical plots in Fig. \ref{fig} show the influence of $\gamma $ and $B$ on entanglement and MINs. The points ($J_{c_1}, J_{c_2}$) are calculated from (\ref{critical pnts}) as (-2.927, -1.268), (-1.163, -0.854), (-0.008, 0.062), (-1.724, -1.102) respectively, corresponding to the plots clockwise starting from top left. These points within which the spins are unentangled as shown in the insets. Here again the non zero MINs in these windows show the presence of correlation between the spins without entanglement. The zero entanglement window (between $J_{c_1} $ and $J_{c_2} $) is also narrowed by the anisotropic parameter $\gamma$ and/or magnetic field $B$. 
\\
\begin{figure*}[!ht]
\centering\includegraphics[width=0.7\textwidth]{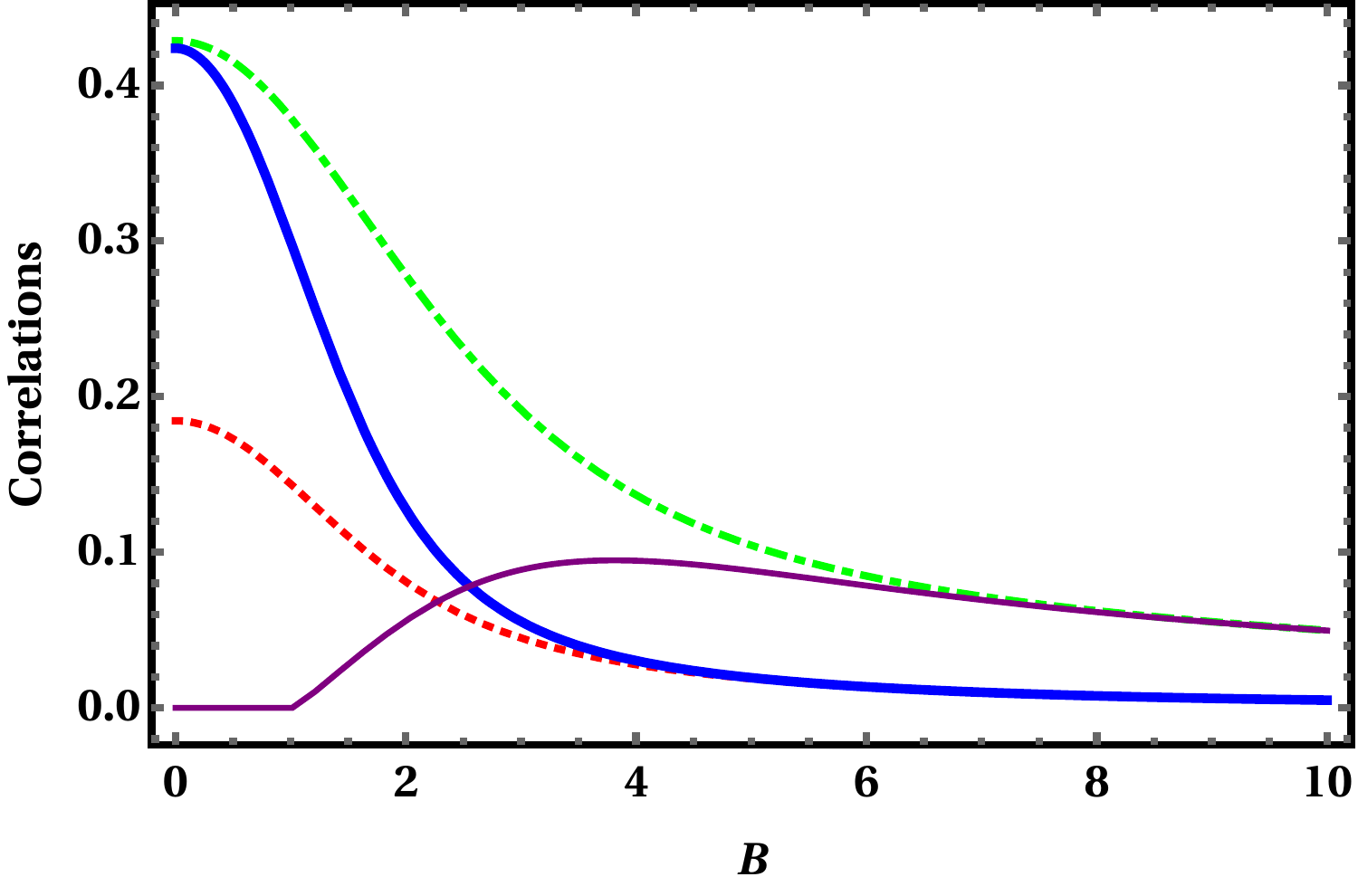}
\caption{(color online) Concurrence (thin), MIN (dotted), T-MIN (dot-dashed) and F-MIN (bold) for XYZ model with $J_z = -1, \gamma = 0.5$ and $J = 2$.}
\label{fiB}
\end{figure*}
\\
In what follows we show the influence of magnetic field $B$ on the correlation measures. Intuitively, one may expect that the intervention of external field decreases entanglement and so the correlation, as shown earlier. Interestingly, for certain choice of parameters we have observations which are different from the earlier one. A typical example is shown in Fig. \ref{fiB}. Here we observe non-zero MINs and zero concurrence for a range of magnetic field, once again showing the existence of spin correlation without entanglement. We also observe that as the field increases the entanglement increases and then decreases.  This observation indicates the possibility of field induced entanglement in spin system. On the other hand, all the three MINs are shown to decrease with the increase of field. Thus the intervension of fields always decreases the quantum signature of the spins as capured by MINs and not by the entanglement. Finally, we also observe that the entanglement and MINs are not significantly influenced by inhomogenity in the magnetic field.
%%%%%%%%%%%%%%%%%%%%%%%%%%%%%%%%%%%%%%%%%%%%%%%%%%%%%%%%%%%%%%%%%%%%%%%%%%%%%%%%%%%%%%%%%%%%%%%%%%%%%%%%%%%%%
\section{Conclusion}
In this paper, we have studied in detail the entanglement of Heisenberg $XYZ$ spin model in presence of external magnetic field at thermal equilibrium. In particular, we have shown the region where the spins are unentangled. Since quantum correlation or nonlocality is much more than entanglement, we quantify the former using different forms of measurement-induced nonlocality (MIN). The MIN, trace MIN and fidelity MIN are shown to be consistent in capturing the correlation. We have also shown the existence of correlation between the spins even in the absence of entanglement. However, the states with maximum entanglement is shown to possess maximum correlation. Finally, we have demonstrated that the correlation of spins is always decreased with the intervention of magnetic field, unlike entanglement. It would be interesting in future to extend the present study in a chain of spins. 
%%%%%%%%%%%%%%%%%%%%%%%%%%%%%%%%%%%%%%%%%%%%%%%%%%%%%%%%%%%%%%%%%%%%%%%%%%%%%%%%%%%%%%%%%%%%%%%%%%%%%%%%%%%%%

\bibliographystyle{99}  

\end{document}